\documentclass[prd, preprint, aps]{revtex4}

\usepackage{graphicx}
\usepackage{color}
\usepackage{amsfonts}

\def\be{\begin{equation}}
\def\ee{\end{equation}}
\def\ba{\begin{eqnarray}}
\def\ea{\end{eqnarray}}
\def\beq{\begin{eqnarray}}
\def\eeq{\end{eqnarray}  }

\def\eref#1{Eq.~(\ref{#1})}

\def\rmd{{\rm d}}
\def\rmO{{\rm O}}
\def\ADM{{\rm ADM}}

\def\BE{{E_b}}
\def\BENorm{{E_b/(M_{AH1}+M_{AH2})}}
\def\lp{\left(}
\def\rp{\right)}

\begin{document}

\newcommand{\myL}{\mathcal{L}}
\newcommand{\del}{\nabla}
\newcommand{\Real}{\mathbb{R}}
\newcommand{\interp}{I_{2h}^h\,}
\newcommand{\prolong}{I_{2h}^h\,}
\newcommand{\restrict}{I_{h}^{2h}\,}
\newcommand{\mkeq}[1]{\begin{equation}#1\end{equation}}
\newcommand{\sep}{d}

\title{Spin Dependence in Computational Black Hole Data}

\author{Scott H. Hawley}
\author{Michael J. Vitalo}
\author{Richard A. Matzner}
\affiliation{Center for Relativity, University of Texas at Austin,
Austin, TX 78712-1081, USA}

\begin{abstract} 

We have implemented a parallel multigrid solver, to solve the
initial data problem for $3+1$ General Relativity. This involves solution of
elliptic equations derived from the Hamiltonian and the momentum
constraints. We use the conformal transverse-traceless method of York and
collaborators~\cite{YP,MY,York,Mathews,Bowen+York} which consists of a conformal
decomposition with a scalar $\phi$ that adjusts the metric, and a vector
potential $w^i$ that adjusts the longitudinal components of the extrinsic
curvature. The constraint equations are then solved for these quantities
$\phi$, $w^i$ such that the complete solution fully satisfies the
constraints. We apply this technique to compare with theoretical expectations for
the spin-orientation- and separation-dependence in the case of spinning
interacting (but not orbiting) black holes.  We write out a formula for the
effect of the spin-spin interaction which includes a result of Wald \cite{WaldPRD}
as well as
additional effect due to the rotation of the mass quadrupole moment of a spinning 
black hole.  A subset of these spin-spin effects are confirmed via our numerical
calculations, however due to computer time limitations the full parameter space
has not yet been surveyed and confirmed.  In particular, at the relatively small separations 
($d \leq 18m$) we are able to
consider, we are unable to confirm the expected asymptotic fall-off of $d^{-3}$ for these
effects.
\end{abstract}

\pacs{ }

\maketitle

\vspace{0.5cm}


\section{Introduction}
\label{sec:intro}
Simulation of binary black hole mergers will play an important part in the prediction,
detection, and the analysis of signals in gravitational
wave detectors. In the usual approach to computing the merger of black holes
(generically called the $3+1$ method), one has first to initiate the
simulation by producing consistent data. Four of the components of the
Einstein equation do not contain time derivatives of the spatial metric, nor
of the momentum of the 3-metric. These components, $G_{00}=0$ and
$G_{i0}=0$, 
are thus called constraint equations, and they must be satisfied in any
specification of initial data. (We are interested in black hole
interactions, which are vacuum, i.e. matter-free, so the right side of the
Einstein equation is zero: $G_{\mu \nu}=0$.) As we recall in the next
section, a conformal decomposition
\cite{YP,MY,York,Mathews,Bowen+York,Choquet,Lichnerowicz} allows the
solution of these components to be put in the form of a set of four coupled
elliptic equations. These elliptic equations are the subject of our work. We
solve them via a multigrid method which applies concepts from
\cite{Hawley+Matzner} to
this problem. We demonstrate the accuracy of our data by considering
features discussed analytically by Wald \cite{WaldPRD}. Wald described the
spin-spin effect on the binding energy of two black holes in an analytic
perturbation scheme, where one hole is much more massive than the other.
Our computational technology is well suited to
simulating these effects for equal mass black holes, and we demonstrate
agreement in some aspects of the
computational spin-spin interactions with the analytic estimate, 
for separations that are not small.

\section{$3+1$ Formulation of Einstein Equations}

We take a Cauchy formulation
(3+1) of the ADM type, after Arnowitt, Deser, and
Misner~\cite{ADM}. In such a method the 3-metric $g_{ij}$ and its momentum
$K_{ij}$ are specified at one initial time on
a spacelike hypersurface, and evolved into the future.  The ADM metric is
\be
\rmd s^2 = -(\alpha^2 - \beta_i \beta^i)\,\rmd t^2 + 2\beta_i \, \rmd t
\,\rmd x^i
     + g_{ij}\, \rmd x^i\, \rmd x^j
\label{eq:admMetric}
\ee
where $\alpha$ is the lapse function and $\beta^i$ is the shift
3-vector; these gauge
functions encode the
coordinatization.\renewcommand{\thefootnote}{\fnsymbol{footnote}}\setcounter
{footnote}{2}\fnsymbol{footnote}

\footnotetext[2]{Latin indices run $1,2,3$ and are lowered and raised
by $ g_{ij}$ and its 3-d inverse $ g^{ij}$. The time derivative will be
denoted by an overdot ($\dot{}$)}

The Einstein field equations contain both hyperbolic evolution equations
and elliptic constraint equations.
The constraint equations for vacuum in the ADM decomposition are:
\beq
H = \frac{1}{2} [R - K_{ij}K^{ij} + K^2] &=& 0,
\label{eq:constraintH}
\eeq
\beq
H^i = \nabla_j \lp K^{ij} - g^{ij}K\rp  &=& 0.
\label{eq:constraintK}
\eeq
Eq. (\ref{eq:constraintH}) is known as the Hamiltonian constraint;
Eq. (\ref{eq:constraintK}) is the momentum constraint (three components).
Here $R$ is the 3-d Ricci scalar constructed from the 3-metric, and
$\nabla_j$ is the torsion-free
3-d covariant derivative compatible with $ g_{ij}$.
Initial data must satisfy these constraint
equations; one may not freely specify all components of $g_{ij}$ and
$K_{ij}$.

One of the evolution equations from the
Einstein system is

\begin{equation}
       \dot g_{ij} = -2\alpha K_{ij} +\nabla_j\beta_{i} + \nabla_i\beta_{j},
\label{eq:gdot}
\end{equation}
and this will prove useful in our data setting procedure below.

\section{Data Form}
\label{sec:DataForm}

Solutions of the initial value problem have been addressed in the past by
several groups,
~\cite{YP,MY,York,Mathews,Bowen+York,Choquet,Lichnerowicz,Hawley+Matzner},
~\cite{Cook,Pfeiffer,Baumgarte,GGB1,GGB2,GGB3,Cook2,Pfeiffer2,CookReview,Shoemaker,Huq}.
It is the case that 
until recently, most data have been constructed assuming that the initial
3-space is conformally flat. The method most commonly used is the approach
of Bowen and York~\cite{Bowen+York}, which chooses maximal spatial
hypersurfaces (for which the quantity $K \equiv {K^a}_a =0$), as well as
taking the spatial 3-metric to be conformally flat.

The chief advantage of the maximal spatial hypersurface approach is
numerical simplicity, as this choice  decouples the Hamiltonian constraint
from the momentum constraint equations.  Besides, for  $K = 0$, if the
conformal background is flat Euclidean 3-space, then there are known
$K_{ij}$ that analytically solve the momentum constraint~\cite{Bowen+York}.
The constraints then reduce to one elliptic equation for the conformal
factor $\phi$.
Very recently substantial success has been achieved evolving
Bowen-York data using ``puncture'' methods \cite{Brownsville,Goddard}.
However, we generally use an alternative choice of background
3-metric, which is based on a metric constructed from single black hole Kerr
Schild data\cite{KerrSchild}; multiple black holes are constructed by a
superposition in the conformal background. It has been shown that this
process, while not exact for multiple black hole data, does contain much of
the physics. It clearly {\it is} exact for a
single black hole, even a spinning or
boosted black hole~\cite{Matzner:1999pt}.

\subsection{Kerr Schild Black Holes}
\label{subsec:KSform}

The Kerr-Schild~\cite{KerrSchild} form of a black hole solution describes
the spacetime of a single black hole with mass, $m$, and specific angular
momentum, $a = j/m$, in a coordinate system that is well behaved at the
black hole horizon:
\be
        \rmd s^{2} = \eta_{\mu \nu}\,\rmd x^{\mu}\, \rmd x^{\nu}
                 + 2H(x^{\alpha}) l_{\mu} l_{\nu}\,\rmd x^{\mu}\,\rmd
x^{\nu},
        \label{eq:1}
\ee
where $\eta_{\mu \nu}$ is the metric of flat space, $H$ is a scalar
function of $x^\mu$, and $l_{\mu}$ is an (ingoing) null vector, null
with respect to both the flat metric and the full metric,
\be
\eta^{\mu \nu} l_{\mu} l_{\nu} = g^{\mu \nu} l_{\mu} l_{\nu} = 0.
\label{eq:2}
\ee

Comparing the Kerr-Schild metric with the ADM
decomposition~\eref{eq:admMetric}, we find that the $t=\hbox{\rm constant}$
3-space metric is: $g_{ij} = \delta_{ij} + 2 H l_i l_j$.
Further, by comparison to the ADM form, we have
\be
\beta_i = 2 H l_0 l_i,
\label{eq:beta_ks}
\ee
and
\be
\alpha = \frac{1}{\sqrt{1 + 2 H l_0^2}}.
\ee
Explicit forms of $H(x^\mu)$ and $l_\alpha(x^\nu)$ for Kerr black holes 
are given in a number of references. 
See \cite{KerrSchild},\cite{Matzner:1999pt},\cite{Marronetti}. 
Many details of the algebraic manipulation of the Kerr-Schild 
form are found in reference \cite{Huq}.

The extrinsic curvature can be computed from Eq.(\ref{eq:gdot}):

\be
        K_{ij} = \frac{1}{2\alpha}[\nabla_j\beta_{i} + \nabla_i\beta_{j}
                     - \dot g_{ij}],
\label{eq:k_ks}
\ee
Each term on the right hand side of this equation is known analytically; in
particular, for a black hole at rest, $\dot g_{ij}=0$.

\subsection{Boosted Kerr-Schild black holes}

The Kerr-Schild metric is form-invariant under a
boost, making it an ideal metric to describe moving
black holes.  A constant Lorentz transformation
(the boost velocity, ${\bf v}$, is specified with respect to the background
Minkowski spacetime) $\Lambda^{\alpha}{}_{\beta}$ leaves the
4-metric in Kerr-Schild form, with $H$ and $l_{\mu}$
transformed in the usual manner:\\
\ba
  x'^{\beta} &=& \Lambda^\beta{}_\alpha x^{\alpha},\\
  H'(x'^{\alpha})  &=&  H\lp (\Lambda^{-1})^\alpha{}_\beta
                          \,\,x'^{\beta}\rp,\\
  l'_{\delta}(x'^{\alpha}) &=& \Lambda^{\gamma}{}_{\delta}\,\,
            l_{\gamma}\lp(\Lambda^{-1})^\alpha{}_\beta\,\, x'^{\beta}\rp .
\label{eq:ks_boost}
\ea
Note that $l'_{0}$ is no longer unity. As the initial solution
is stationary, the only time dependence comes in the
motion of the center, and the full metric is stationary with a Killing
vector reflecting the boost velocity.
The boosted Kerr-Schild data exactly represent a spinning and/or moving single
black hole. 

\subsection{Background data for multiple black holes}

The structure of the Kerr-Schild metric suggests a natural extension
to generate the background data for multiple black hole spacetimes.
We first choose mass and angular momentum parameters for each hole,
and compute the respective $H$ and $l^\alpha$ in the appropriate
rest frame.  These quantities are then boosted in the desired direction
and offset to the chosen position in the computational frame.
The computational grid is the center of momentum frame for the two holes,
making the velocity of the second hole a function of the two
masses and the velocity of the first hole.
We compute the
individual metrics and extrinsic curvatures in the coordinate system
of the computational domain:
\beq
   {}_A g_{ij} &=&  \eta_{ij}
                     + 2~{}_A H ~{}_A l_{i} ~{}_A l_{j},\\
   {}_A K_i{}^m &=& \frac{1}{2\alpha} ~{}_A g^{mj}
               \lp \nabla_j ~{}_A\beta_{i} + \nabla_i~{}_A \beta_{j}
                    - ~{}_A \dot g_{ij}\rp.
\eeq
The pre-index $A$ labels the black holes.
Background data for $N$ holes are then constructed in superposition:
\beq
\tilde{g}_{ij} &=& \eta_{ij} + \sum_A^N 2~{}_A H {}_A l_i ~{}_A l_j ,\\
\tilde{K} &=& \sum_A^N ~{}_AK_i{}^i ,\\
\tilde{A}_{ij} &=& \tilde{g}_{n(i}~~\sum_A^N \lp {}_AK_{j)}{}^n
                  -\frac{1}{3} \delta_{j)}{}^n ~{}_AK_i{}^i\rp .
\label{eq:ks_super}
\eeq
A tilde ( $\tilde{}$ ) indicates a background field tensor. Notice 
that we do {\it not} use the attenuation functions introduced by 
Bonning et al.\cite{Bonning}.

\begin{figure}
\begin{center}
\includegraphics[width=6.0in, angle=0]{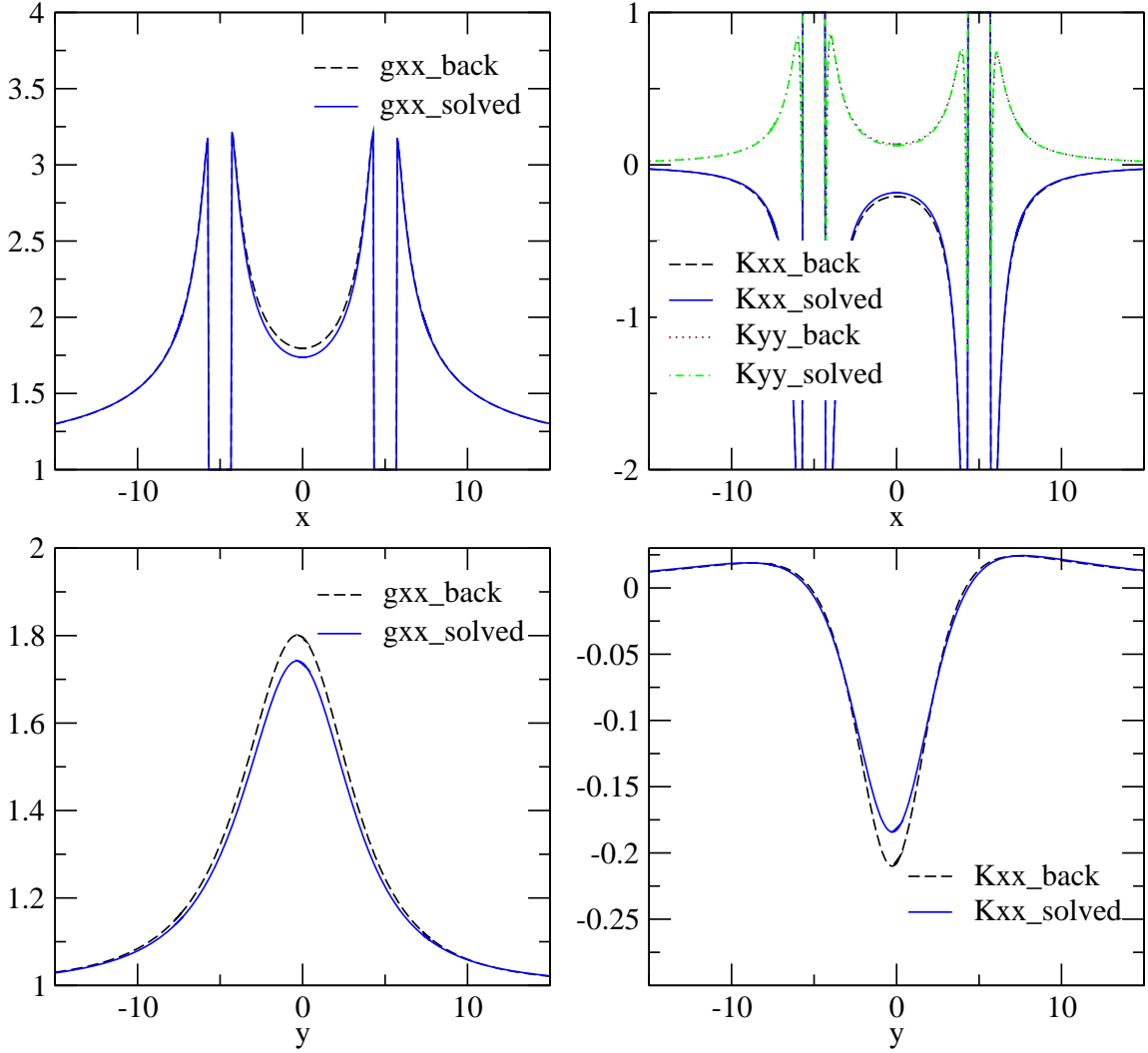}
\end{center}
\vspace{-1.0cm}
\caption{
Comparison of background Kerr-Schild superposition data (dashed lines) 
with the final output of our elliptic constraint equation solver (solid
lines).
We see that the background is quite close to the physical solution.
These particular data were generated for two holes located at $x=\pm 5m$, with
spins $a_1=a_2=0.5$, with the spin of the $x=-5m$ hole tipped by rotation about the $x$ axis by $\theta_1=7\pi / 8$,
and an excision radius of $0.75m$.
}
\label{fig_comp_back_out}
\end{figure}

To give the reader a feel for how closely the Kerr-Schild superposition
data resemble a true binary black hole spacetime, in Figure \ref{fig_comp_back_out}
we provide a graph
comparing the superposed Kerr-Schild background data with
the subsequent solutions of the constraint equations (described below).

\section{Generating the physical spacetime}

We will consider in this paper physical applications which use
superposed Kerr-Schild backgrounds. When multiple black holes are
present, the background superposed Kerr-Schild data described in
the previous section are not solutions of the constraints,
Eqs.~(\ref{eq:constraintH})--(\ref{eq:constraintK}). Hence they do
not constitute  a physically consistent data set. A physical spacetime
can be constructed by modifying the background fields with new
functions such that the constraints {\it are} satisfied. We adopt
the conformal transverse-traceless method of York and
collaborators~\cite{YP} which consists of a conformal decomposition
and a vector potential that adjusts the longitudinal components of
the extrinsic curvature. The constraint equations are then solved
for these new quantities such that the complete solution fully
satisfies the constraints.  We do not consider ${\rm tr}K=0$, nor
conformally flat, solutions.

The physical metric, $g_{ij}$, and the trace-free part of the extrinsic
curvature, $A_{ij}$, are related to the background fields through a
conformal
factor
\ba
g_{ij} &=& \phi^{4} \tilde{g}_{ij}, \label{confg1} \\
A^{ij} &=& \phi^{-10} (\tilde{A}^{ij} + \tilde{(lw)}^{ij}),
\label{eq:conf_field}
\ea
where $\phi$ is the conformal factor, and $\tilde{(lw)}^{ij}$
will be used to cancel any possible longitudinal contribution to the
superposed background extrinsic curvature.
$w^i$ is a vector potential, and
\ba
\tilde{(lw)}^{ij} \equiv \tilde{\nabla}^{i} w^{j} + \tilde{\nabla}^{j} w^{i}
        - \frac{2}{3} \tilde{g}^{ij} \tilde{\nabla_{k}} w^{k}.
\label{lw}
\ea
The trace $K$ is taken to be a given function
\be
K = \tilde K.
\label{tk}
\ee
Writing the Hamiltonian and momentum constraint equations in terms of
the quantities in 
Eqs.~(\ref{confg1})--(\ref{tk}), we obtain four coupled
elliptic equations for the fields $\phi$ and $w^i$~\cite{YP}:
\ba
\tilde{\nabla}^2 \phi &=&  (1/8) \big( \tilde{R}\phi
        + \frac{2}{3} \tilde{K}^{2}\phi^{5} -  \nonumber \\
        & & \phi^{-7} (\tilde{A}{^{ij}} + (\tilde{lw})^{ij})
            (\tilde{A}_{ij} + (\tilde{lw})_{ij}) \big),   \label{ell_eqs1}
\\ 
\tilde{\nabla}_{j}(\tilde{lw})^{ij} &=& \frac{2}{3} \tilde{g}^{ij} \phi^{6}
        \tilde{\nabla}_{j} K - \tilde{\nabla}_{j} \tilde{A}{^{ij}}.
\label{ell_eqs}
\ea

Our outer boundary condition for $\phi$, namely
\begin{equation}
 \partial_{\rho} \left( \rho (\phi - 1) \right)|_{\rho \rightarrow
\infty} = 0.
\label{eq:phi_boundary}
\end{equation} 
enforces $\phi \rightarrow 1 $ at $\infty$, but does not specify the size
(or sign) of the $\rho^{-1}$ term in $\phi$. (Here $\rho^2 = x^2+y^2+z^2$.) We also take as boundary
conditions for the vector $w^i$:
\begin{eqnarray}
   &  & \partial_\rho (\rho w^{i} n_{i}) = 0, 
\label{25} \\[.12in]
   &  & \partial_\rho \left( \rho^{2} w^{i} (\delta_{ij} - n_{i}n_{j})
   \right) = 0\,,
\label{26}
\end{eqnarray}
where $n_i$ is the outward pointing unit spatial normal.
Condition (\ref{eq:phi_boundary}) is a {\it Robin} condition
commonly used for computational conformal factor determination.
Conditions (\ref{25}) and (\ref{26}) were derived by
Bonning et al.\cite{Bonning}.

\section{Numerical Methods}

We first discuss the computational code and tests, and
some code limitations.

The constraint equations \eref{ell_eqs1}, \eref{ell_eqs} are solved
with a
multigrid solver~\cite{Hawley+Matzner}.
The present code is essentially the same as that described in
\cite{Hawley+Matzner},
except that it has been extended to the full set of constraint equations,
non-flat backgrounds, and features parallel processing.
The multigrid scheme is essentially a clever means of
eliminating successive wavelength-components of the error via the
use of relaxation at multiple spatial scales.  It makes use of
some sort of local averaging procedure (e.g. Gauss-Seidel relaxation).
Such relaxation is extremely
effective at eliminating short-wavelength components of the error,
or in other words, at ``smoothing'' the error (i.e.,  the residual,
see below). However, relaxation fails to operate efficiently
on long-wavelength components of the error (components that involve
discretization points more than a few away from the
point at which the solution is sought).
Multigrid addresses the solution repeatedly on grids of different
discretization, achieving the same efficiency at smoothing every scale.

Because the implementation in the method is also in \cite{Hawley+Matzner}, 
we do not repeat it here.
\section{Multigrid with Excised regions}

In our formulation, the black holes are represented by excised regions.  Because
we work in
Cartesian coordinates, and because we want completely general
implementation, 
we do not typically expect that the excision will be defined by
overlapping points on the various grids of different resolution.

\begin{figure}
\centering
\includegraphics[width=5.5in]{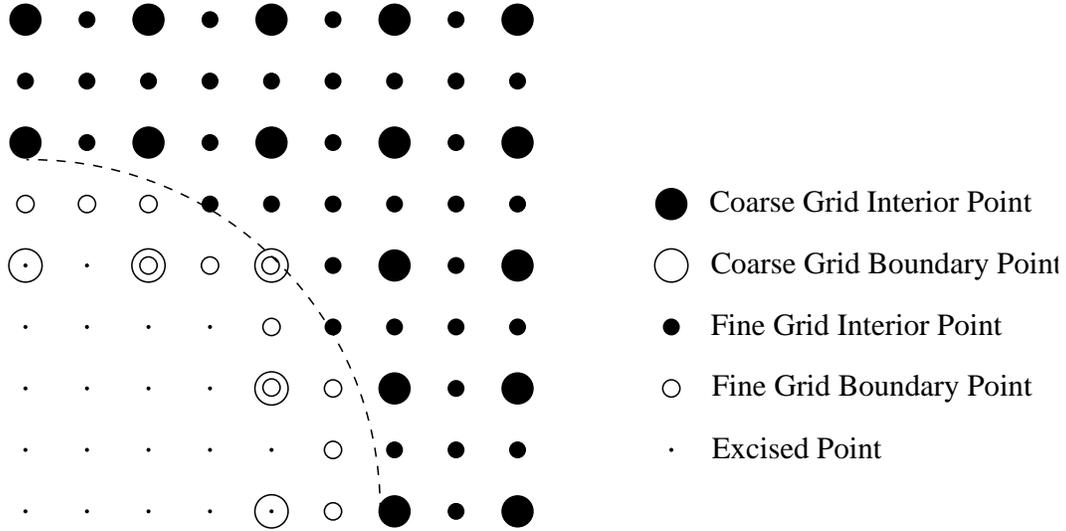}
\caption{
Example of how the inner boundary is defined, showing points on a coarse
grid and a fine grid. Inner boundary points are those points which are
immediately interior to a circle of radius $r_{\rm ex}$. The large filled
circles show normal interior grid points (i.e., non-excised, non-boundary
points) on the coarse grid, and the large open circles show boundary points
on the coarse grid.  The small filled and open circles show fine grid
interior points and boundary points, respectively. The small dots show
excised points on the fine and/or coarse grids, as appropriate. (Only one
quadrant of a full domain is shown in this picture for purposes of clarity.)
}
\label{fig_define_ex_reg}
\end{figure}

Our definition of the excision region is that on each grid, the inner
boundary consists of points that lie {\it just inside}, i.e. up
to one grid point inside,  the analytic location of the inner boundary, as shown in
Figure \ref{fig_define_ex_reg}.
While there are exceptional configurations such as cubic excision defined
so that the excision boundary lies on points of the coarsest grid, this
definition means that generically the size of the excision is {\it larger}
on 
the finer grids.  

\begin{figure}
\centering
\centerline{\includegraphics[width=5.2in,height=1.23in]{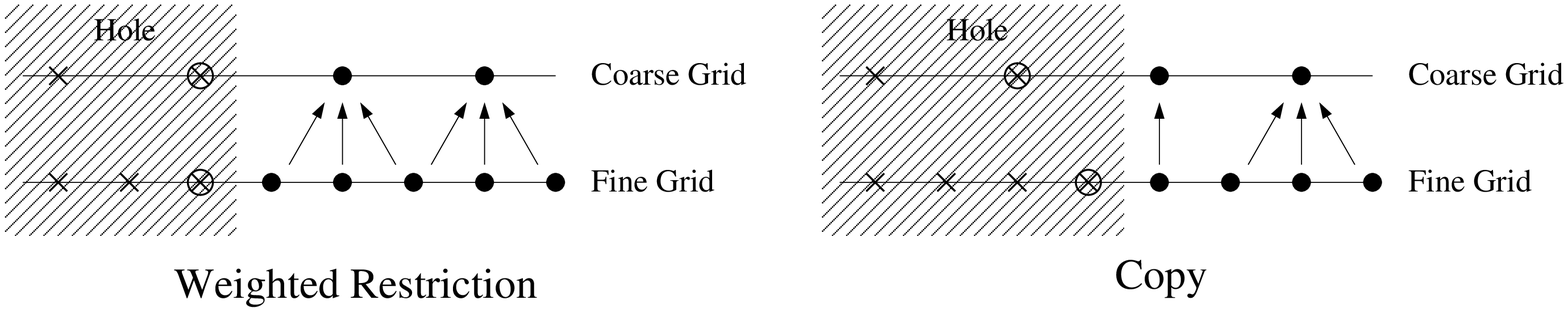}
}
\caption{1-D schematic of scheme for restriction scheme near inner 
boundary.
The circles on the rightmost X's indicate that this is where the
Dirichlet conditions are applied, i.e., there {\em are} data on these
points. 
One {\em could} use these points in weighted restriction even
in the case shown in the right panel.  However, we choose never to use
these boundary data in weighted restriction, and instead do a
simple ``copy" operation.  The boundary points themselves are 
updated using either a direct copy,
or for coarse grid points over excised fine grid points
via an average over parabolic fits of 
fine-grid data in all available directions. }
\label{fig_restrict_scheme}
\end{figure}

This definition of the inner boundary affects the way in which data are
restricted from fine grids to coarse grids.  Away from the inner boundary,
weighted restriction is performed, as shown in left pane of Figure 
\ref{fig_restrict_scheme}.  However, if any of the points used in the 
weighted average lie on an inner boundary, then these points are not used
and instead a simple ``copy" operation is performed as shown in the right pane
of Figure \ref{fig_restrict_scheme}.   The inner boundary points
themselves may need to be filled in on coarse grids (since on the
fine grid they may be excised), and to do this we apply a weighted
``inward extrapolation'' using a parabolic fit to surrounding fine
grid points.

This scheme has been implemented in a parallel computing environment, using
{\it MPI} to communicate between processors. Each processor handles a part 
(a {\it patch}) of the total domain. The patch is also logically surrounded by  
``ghost zones". Because we deal with a finite-difference representation of
derivatives, the communication between processors requires the filling of
these ``ghost zones" on the borders of the patches, using values computed on
other processors, so that derivatives can be accurately computed near the
boundaries of the patches. This has implications for the way that smoothing
is handled in our simulation.

On a single processor Gauss-Seidel smoothing proceeds across the grid, and
the updates at any particular point involve some surrounding points that
have been updated and some that have not been. If the same scheme has been
implemented on two processors (say splitting the $x-$axis), the buffer
region of one patch will already have been updated when the smoother of the
other patch begins to use the equivalent points. The order and direction of
the filling of the ghost zones can lead to inconsistent behavior ({\it i.e.,} the result
will be different from the single processor result). One solution to this is
to insert ``wait" commands into the parallel code, so that processors wait
to carry out the process in the correct order. This has the effect of
slowing the execution, and loses the advantage of parallel processing. A
better approach is to use something like {\it red-black} Gauss-Seidel. (In
2-d the red-black pattern is like that on a checkerboard.) If the
differential operator involves only diagonal second derivatives  (no mixed
partials) then each point is updated using only points of the opposite
color. Then all the reds can be updated before any of the blacks, and vice
versa. This ameliorates the ghost zone synchronization problem; the ghost
zones can be maintained in the correct state for every step. In this case
parallelization works as anticipated.

If the background is taken as flat space, then these conditions apply. But
we work with Kerr-Schild
forms of the metric which guarantee that there will be mixed partial
derivatives 
in the operators, and the parallel synchronization problem reappears.

Our solution is to introduce what we call {\it rainbow} smoothing, in which
we make a total of {\it eight} passes (like the two passes in red-black smoothing)
over the grid, where each pass has a stride width of two over each of the
three dimensions of the grid.

\section{Verification of constraint Solution}
\label{sec:correctness}

To verify the solution of the discrete equations, we have examined the
code's convergence in some detail.  We use a set of
completely independent ``residual evaluators"
for the full Einstein system (here applied only to the initial data),
originally constructed
by Anderson \cite{mattDiss} .
These evaluate the
Einstein tensor, working just from the metric produced by the
computational solution, to return fourth order accurate
results. They are completely different from the way
the equations are expressed in the constraint solver code.

\begin{figure}
\begin{center}
\includegraphics[width=6.0in, angle=0]{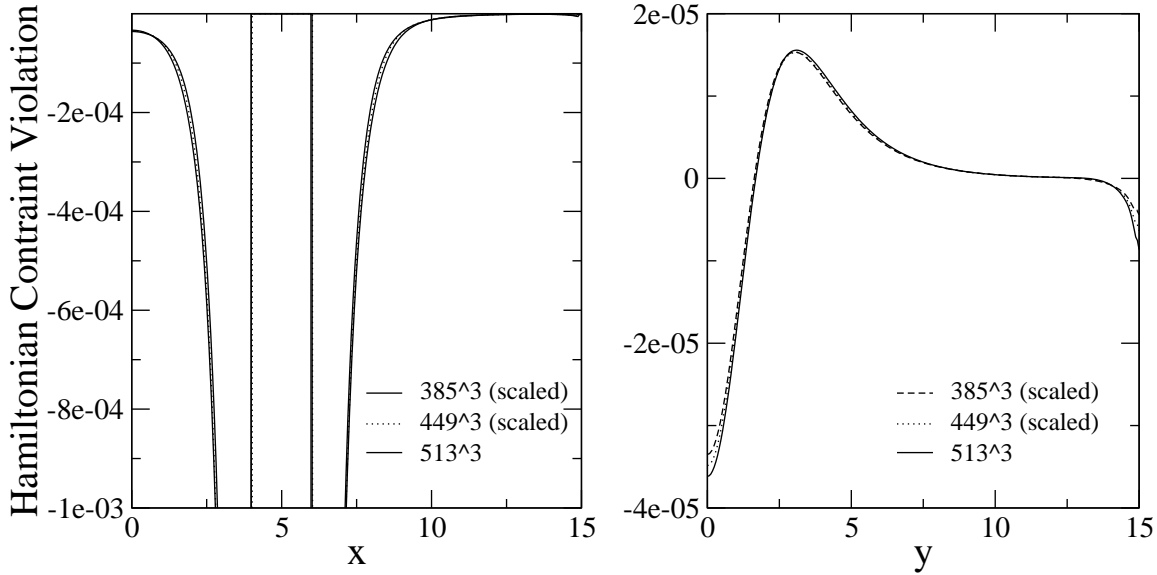}
\end{center}
\vspace{-1.0cm}
\caption{
Convergence of the Hamiltonian constraint along the positive $x$ and $y$
axes.
We show values of the constraint obtained and three different finest-grid
resolutions, $385^3$, $449^3$ and $513^3$, scaled by appropriate ratios of
the mesh spacings consistent with second-order convergence.  We see that
there
is good convergence everywhere except near the outer boundaries.  Because of
this loss of convergence near the outer boundaries, we evaluate the ADM mass
over the surface a cube with half-width 12M.  (In the left pane,
the vertical scale has been exaggerated in order to zoom in on the 
``body'' of the domain, and cuts out the peaks immediately adjacent to
the excision regions.
}
\label{fig:ham_conv}
\end{figure}

Figure \ref{fig:ham_conv} shows
such a plot of convergence for 
the Hamiltonian constraint in an equal mass binary black hole spacetime.
The holes are located at $\pm 5m$ on the $x$-axis, where $m$ is the mass of
one hole. 
The elliptic equations were then solved on grids of sizes $385^3,$
$449^3,$ and $513^3,$ giving finest-grid resolutions of approximately
$m/12.8$, $m/15$, and $m/17$. We use a five-level multigrid hierarchy.
Figure \ref{fig:ham_conv} demonstrates almost perfect second order convergence, except
near the outer 
boundary, where the convergence is apparently first order. 
The second order convergence shows
that we have achieved a correct finite difference solution to the initial data
problem.

\section{Computational
Limitations on grid coarseness in excised black hole spacetimes}

In the examples given here, where we
work on a fixed spatial domain (say $\pm 15m$), the finest resolutions of
$385^3$, $449^3$, $513^3$, used in the convergence test correspond to
coarsest-grid resolutions of $25^3$, $29^3$, $33^3$, respectively. For the $33^3$
grid, the $\pm15m$ domain is discretized at about $1m$ resolution.

The problem of required resolution for black hole simulations has been
discussed in this context at least since the early {\it Grand Challenge}\cite{BBHGC}
efforts. Present computational resources allow much larger grid size than in
the Grand Challenge epoch, so the conflict appears at higher resolutions
and larger physical domains than previously, and we can do substantial
physics with the present configuration. Our approach will be to introduce a
multiresolution scheme to maintain required resolution near the central
``action", and allow coarsening further away. To accomplish this we are
investigating a mesh-refined multigrid, similar to that described by Brown 
and Lowe\cite{brown}.
However, for the present work, we simply use very high resolution, the
highest that we can presently achieve on the computers available to us,
namely $513^3$ points using 32 processors.

\section{Spin-Spin Effects in Black Hole Interaction}

Wald \cite{WaldPRD} directly computes the force for stationary sources with
arbitrarily oriented spins. He considered a small black hole as a
perturbation in the field of a large hole. The result found
for the spin-spin contribution to the binding energy is

\be
\BE = - \lp \frac{ \vec{S} \cdot \vec{S'} -
        3(\vec{S} \cdot \hat{n})(\vec{S'} \cdot \hat{n})}{\sep\,^3} \rp.
\label{BEWald}
\ee

\noindent Here, $\vec{S}$, $\vec{S}'$ are the spin vectors of the
sources and $\hat{n}$ is the unit vector connecting the two sources,
and $\sep$ is any reasonable measure of separation that approaches
the Euclidean distance $\times (1+\rmO(\sep^{-1}))$ at large $\sep$
(such as the distance measured in the flat background used in the
initial data setting).
Dain~\cite{Dain}, using a definition of intrinsic mass that differs
from ours (see below), finds binding energy which agrees with Wald's
(\eref{BEWald}) at $\rmO(\sep^{-3})$.

\subsection{Computational Spin-Spin Effects in Black Hole Binding Energy}

In order to investigate a computational implementation validating
\eref{BEWald}, we begin with a standard definition of the binding
energy for black hole interactions.

The total gravitational energy in a binary system can be computed
from the initial data using the ADM mass $M_{\ADM}$, which is
evaluated by a distant surface integral (see \eref{eq:adm_mass} below), and
gives 
the Newtonian gravitational mass as measured ``at infinity". For
a measure of each hole's intrinsic mass, we use the horizon mass $M_{\rm
AH}$
defined by (\eref{eq:mirr}) below. Thus the binding energy, $\BE$, is defined as

\be
\BE = M_{\ADM} - M_{\rm AH} - M'_{\rm AH}. \label{binding}
\ee

The ADM mass is evaluated in an asymptotically flat region
surrounding the system of interest, and in Cartesian coordinates is given by
\beq
M_{\ADM} &=& \frac{1}{16\pi} \oint \left( \frac{\partial g_{ji}}{\partial
  x^{j}} - \frac{\partial g_{jj}}{\partial x^{i}} \right)
  \rmd S^i,
\label{eq:adm_mass}
\eeq

The apparent horizon is the only structure available to measure the
intrinsic mass of a black hole.\renewcommand{\thefootnote}{\fnsymbol{footnote}}\setcounter
{footnote}{3}\fnsymbol{footnote}
\footnotetext[3]{Dain \cite{Dain} considers black hole slicings 
that have a second asymptotically flat infinity, and measures a 
mass (an intrinsic mass for the black hole) at this second infinity. 
This approach is impossible for the Kerr-Schild data we consider 
because Kerr-Schild slices intersect the black hole singularity.}
  Complicating this issue is the intrinsic
spin of the black hole; the relation is between horizon area and
{\it irreducible} mass:

\be
A_{\rm H} = 16 \pi m_{irr}^2 = 8 \pi m \lp m + \sqrt{(m^2 -a^2)}\rp.
\label{eq:mirr}
\ee
As \eref{eq:mirr} shows, the irreducible mass is a function of both the
mass and the spin, and in general we have no completely unambiguous way
specify the spin of the black holes in interaction. But, as was shown in
\cite{Bonning}, the spin evolves only very little until the black
holes are very close together. Further, the apparent horizon coincides
closely 
with the event horizon unless the black holes have strong interaction.
Hence we assume that the individual spins
are correctly given by the spin parameters ${}_Aa$ specified in forming
the superposed Kerr-Schild  background, and that the mass is that determined
by 
\eref{eq:mirr} using the area determined from the apparent horizon area.

The physical idea in determining the binding energy is that the
configuration is assembled from infinitely separated black holes, 
which are initially on the $x$-axis and which initially have parallel 
spins. (No energy is required to orient the coordinate system or to 
adiabatically rotate the spins while the holes are infinitely separated.) 
Thus these separated holes have their isolated total energy, i.e. $2m$, 
for equal mass black holes. 

Then one of the black
holes is adiabatically brought to a particular distance from the other, 
for instance a coordinate distance of $10m$ as in some of our examples. 
This is the base configuration from which our computations will start. We then
consider the change in the binding energy as we move the direction of the
spin axis of one of the black holes.

\subsection{ADM angular momentum}

Besides the mass, ADM formulae also exist for the momentum $P^{\ADM}_{k}$
and angular momentum $J^{\ADM}_{ab}$. These formul\ae\ are also
evaluated in an asymptotically flat region surrounding the system of
interest~\cite{Wald, note4}:

\beq
   \label{eq:adm_mom}
P^{\ADM}_{k} &=& \frac{1}{8\pi} \oint \left( K_{ki} - K^{b}{}_{b}
\delta_{ki}
   \right)\rmd S^i,\\
   \label{eq:adm_ang_mom}
J^{\ADM}_{ab} &=& \frac{1}{8\pi} \oint \left( x_{a}K_{bi} - x_{b}K_{ai}
   \right) \rmd S^i.
\eeq
In the data we set, the total momentum is set to zero, so $P^{\ADM}_k=0$.
In general, we set data for arbitrarily spinning holes with arbitrary orbital
impact parameter, so in general the angular momentum $J^{\ADM}_{ab}$
is nonzero, and interesting. In the results presented below as 
code tests, we seek initially non-moving black holes, so the total angular 
momentum $J^{\ADM}_{ab}$ is simply the sum of the intrinsic angular momenta, 
$\Sigma {}_A m {}_a a$.

\subsection{Computational Results}

\begin{figure}
\centering
\includegraphics[width=5.75in]{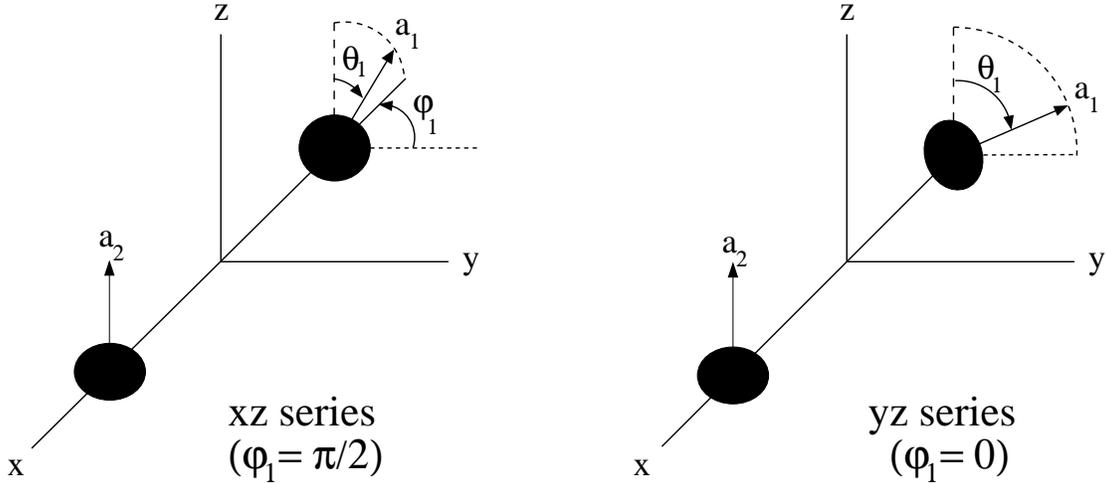}
\caption{The two different BBH configurations investigated. In all cases, the black hole at $x=+d/2$ is held
fixed with a constant spin of $a_2=0.5$ in the $z$ direction.  In the ``$xz$ series," the spin axis of hole 
at $x=-d/2$ is rotated in the $x$-$z$ plane ({\it i.e.} about the $y$ axis) by varying angle $\theta_1$ 
{\it away from} the other hole (holding $\varphi_1=\pi/2$).  
In the ``yz series," this axis is
rotated in the $y$-$z$ plane by varying $\theta_1$ clockwise about the positive $x$ axis.
We note that, as an historical artifact of the background generator code of \cite{Bonning}, angles are defined such
that $\varphi=0$ corresponds to the $y$ axis, {\it not} the (more typical) $x$ axis.
}
\label{fig_xzyz}
\end{figure}

We carried out several series of computational experiments to investigate
the spin-spin interaction. In particular we considered instantaneously
nonmoving black holes of equal
mass $m_1=m_2=m$, with equal spin parameter $a_1=a_2 = 0.5m$. 
The background separation $\sep$ for each series was varied from $6m$ to $18m$.
For instance, we considered $\sep=10m$ 
(holes at coordinate location $x_1= -5m$, $x_2=5m$). 
We varied the spin axis of one hole in
two different planes, resulting in two ``series'' of data.
The hole at $x=+5m$ (``hole number 2'') was maintained with spin $a_2$ aligned with the $z$-axis,
while the direction of the other spin $a_1$ was varied in a plane in 
$\pi/8$ steps through $2 \pi$ from the $+z$-axis through the $-z$-axis and
on back to the $+z$-axis. The difference in the
two series is that in one case  (the ``$yz$ series")
the spin remains in the $y$-$z$ plane; in the other (the ``$xz$ series") it
remains in the $x$-$z$ plane.  These two configurations are displayed in Figure \ref{fig_xzyz}.

The domains we used were typically $\pm 15m$, using $513^3$ grid points, and typically excising
a region of size $r_{ex}=0.9m$.  
The ADM mass was evaluated on a cube with sides at $\pm 12m$ ({\it i.e.}, well inside
``convergence region'' shown in Figure \ref{fig:ham_conv} ).    
(Variations in domain size, resolution, and excision region size
were conducted to estimate the dependence of the resulting binding energy on these physically irrelevant
but computationally important parameters.
For example we conducted a series of runs with outer boundaries
at $\pm20m$ with $513^3$ points, evaluating the ADM mass at $\pm 17m$.)
The apparent horizon areas were
determined using Thornburg's horizon finder \cite{ThornburgAHFinder} in the Cactus 
\cite{cactus-grid, cactus-tools, cactus-review, cactus-webpages,Goodale02a}
computational toolkit, via a post-processing run on our output files.

As shown in Figures 
\ref{be_vs_theta_yz} 
and
\ref{be_vs_theta_xz} 
below, the angular
dependence for the binding energy behavior in the $yz$ case is close to that
predicted by Wald (\eref{BEWald}).

\begin{figure}
\centering
\includegraphics[width=4in]{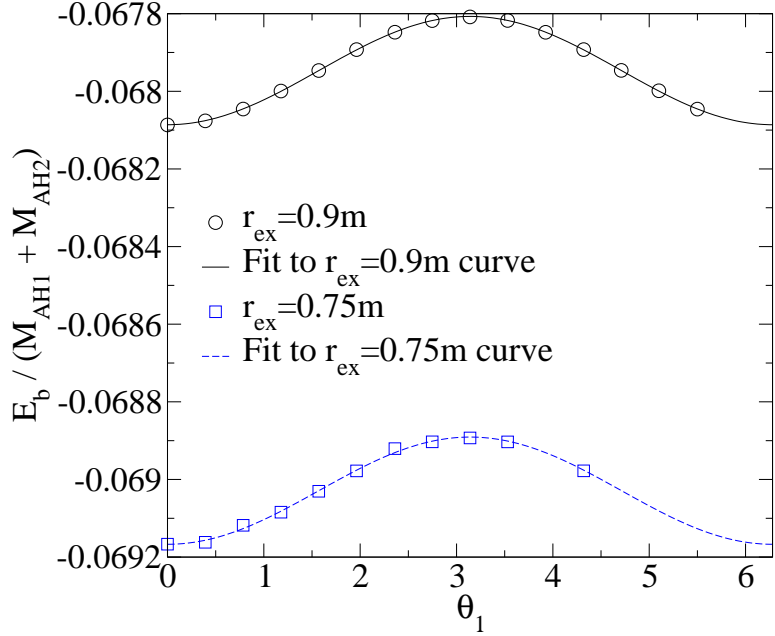}
\vspace{-0.5cm}
\caption{(Normalized) Binding energy vs. spin angle for the $yz$ series.
Present in the graph are two curves corresponding to different 
excision radii $r_{ex}$.  For instance a least-squares fit 
to the $r_{ex}=0.9m$ curve is
$\BENorm = -0.06794 - 1.396\times10^{-4}\cos\theta$.
For the $r_{ex}=0.75m$ curve, the amplitude of the cosine is $1.381\times10^{-4}$.
This cosine corresponds to the $ \vec{S} \cdot \vec{S'}$ term in (\ref{BEWald}).
We note that changing the excision radius changes the overall constant offset of the binding energy,
but does not have a large effect on the amplitude of the spin-spin interaction. 
}
\label{be_vs_theta_yz}
\end{figure}

\begin{figure}
\centering
\includegraphics[width=4in]{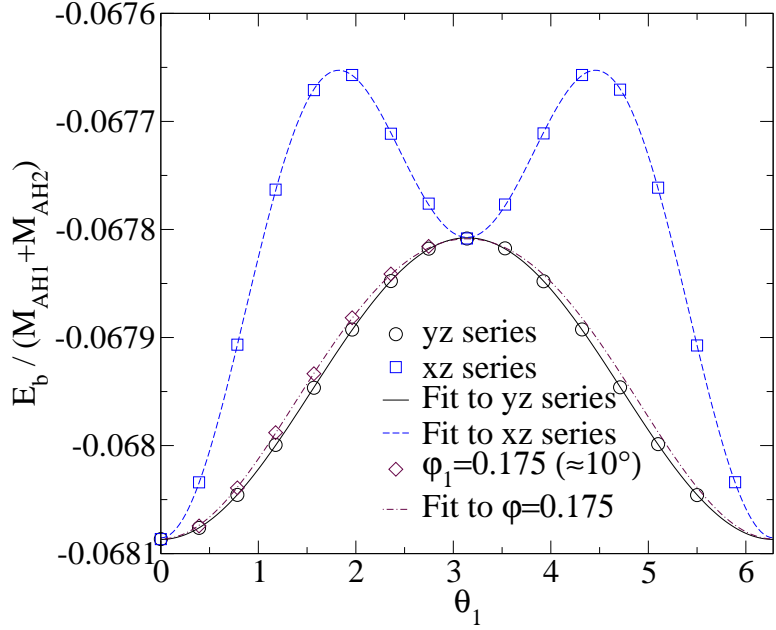}
\vspace{-0.5cm}
\caption{(Normalized) Binding energy vs. spin angle for the $xz$ series
 (``$\varphi_1=\pi/2$''),
along with the $r_{ex}=0.9m$ curve from Figure \ref{be_vs_theta_yz} shown for 
comparison.  Also included is a least-squares fit to the 
$xz$ series:
$\BENorm =  -0.0679 - 1.390\times10^{-4}\cos\theta - 2.762\times10^{-4}\sin^2\theta$.
Note that the coefficient of the $\sin^2\theta$ term is roughly twice
that of those $\cos\theta$ terms both in this graph an Figure \ref{be_vs_theta_yz}.  
Rather than this near factor of 2 we find numerically,
considerations based on the mass quadrupole moment 
would suggest a factor closer to $3/2$, as indicated by (\ref{mass_quad}).
}
\label{be_vs_theta_xz}
\end{figure}

Figure \ref{be_vs_theta_yz} contains two tests of the $yz$ series, computed
identically except for a change in the excision radius. We see that the
spin dependence of the binding energy is unchanged, but there
is an offset in the average binding energy. This binding energy offset 
($0.001m$ out of $-0.07m$) 
is a well known -- but small -- dependence on the inner boundary condition 
in the computation of initial data sets for binary black
hole systems (see, {\it e.g.} \cite{Pfeiffer2}, \cite{Bonning}). It implies an {\it accuracy} in the binding energy
(estimated from the value of the offset)
of less than $2\%$ of the total binding energy. On the other hand, the 
behavior of the spin-dependence,
the cosine curve in the binding energy, implies a {\it precision} much smaller
than the peak-peak amplitude of the cosine curve; we estimate $0.02\%$, 
one tenth of the peak to peak amplitude.

However, Figure \ref{be_vs_theta_xz} for the $xz$ series (where one of the spins tips
{\it away from} the other) reminds us that there are additional physical
effects in play. 
{\it The Kerr solution has a quadrupole moment arising nonlinearly from its spin.} 
In terms of Newtonian potential for Kerr:
\beq
\phi = -\frac{m}{\sep} -\frac{3}{2 \sep ^3} ma^2\cos^2\Theta + ... ;
\label{mass_quad_init}
\eeq
the quadrupole term is the $\cos^2 \Theta$ term,
where $\Theta$ is the ``viewing'' angle at which one hole ``sees'' the other.
Given our configurations in which the spins by default are perpedicular to the line of
separation, $\Theta=\pi/2$ when $\theta=0$.  Thus in terms of our spin-orientation 
angle $\theta$ the effect may be written as
\beq
\phi = -\frac{m}{\sep} -\frac{3}{2 \sep ^3} ma^2\sin^2\theta + ... ;
\label{mass_quad}
\eeq
Here $\sep$ is a radial
coordinate, defined so that angular dependence begins in the metric
only at $O(\sep^{-3})$\cite{KipMoments}. Hernandez \cite{walterHernandez}
expands the asymptotic Kerr-Schild form and comes
to the same result for the quadrupole moment of the Kerr black hole.
The quadrupole $\cos^2\Theta$ effect is not evident in the $yz$ series because
the hole at $+5m$ is always ``looking" at the equator of the hole at $-5m$, i.e. at
$\Theta=\pi/2$ so there is zero effect.
even as the hole at $-5m$ tilts. However, since the $xz$ series tilts the hole at $x=-5m$
away from hole at $x=+5m$, the fixed hole ``sees" different latitudes of the rotated hole
in the $xz$ series. 

Bonning et al.\cite{Bonning} showed that Kerr-Schild data correctly predicts
the Newtonian binding energy $-mm'/{\sep}$. The total binding in a relativistic 
calculation is this $\rmO(\sep^{-1})$ term, plus Wald's $\rmO(\sep^{-3})$ 
spin- spin interaction, plus the quadruole terms in the potential, plus any possible $\rmO(\sep^{-2})$
contribution to the solution.

Following Wald's notation then, the complete spin dependence may be written as

\beq
\BE = - \lp \frac{ \vec{S}_1 \cdot \vec{S}_2 -
        3(\vec{S}_1 \cdot \hat{n})(\vec{S}_2 \cdot \hat{n})}{\sep\,^3} \rp.
      + { 3 \over 2\sep\,^3}  \lp 
              {m_2\over m_1} [\vec{S}_1\cdot\hat{n}]^2 + {m_1\over m_2} [\vec{S}_2\cdot\hat{n}]^2 \rp.
\label{HVMEq}
\eeq
(Note that, for the configurations considered in this paper, $\vec{S}_2 \cdot \hat{n} = 0$.)
Both the Wald spin-spin term and the quadrupole moment term in the
expansion of the potential are proportional to $\sep^{-3}$, though this
is correct only near infinity; for close distances (such as
$\sep \approx 10m$ considered here) one expects deviations from nonlinear
terms in the results. In fact the angular dependence
of these terms is remarkably accurately reproduced.  Table \ref{table_coeffs} shows
the first {\it four} coefficients in fits to the binding energy; the
third and fourth power coefficients are substantially below the cosine
and cosine-squared coefficients. The Wald formula would
produce an amplitude of $0.5^2/10^3 = 2.5 \times 10^{-4}$ for our
case of equal spins of $a=0.5 m$ and separation of $\sep=10m$; the
actual coefficient from the fit (after multiplying by the sum of the horizon masses) is 
$2.97 \times 10^{-4}$.  This apparent agreement is somewhat of an accident, however,
since the expected dependence of $d^{-3}$ is not present in our data, as we will
show below.
The term
arising from the the quadrupole term (the cosine squared term) suggests
a coefficient of $3.75 \times 10^{-4}$  ($1.5$ times the expected
amplitude of the spin-spin term). Our fit to the experiment
(the $xz$ series) produces $5.860\times 10^{-4}$.

The Wald formula, Equation (\ref{BEWald}), predicts no
difference in the cosine term ($A_1$ in Table \ref{table_coeffs}) between the $xz$ and
$yz$ series. (The ($\vec{S}\cdot\hat{n}$) term in Eq. (\ref{BEWald})
is zero for all experiments carried out because the
black hole on the positive $x$ axis has a fixed spin direction
parallel to the $z$ axis.)  This is the behavior we find; compare the coefficients $B_1$ for 
$\cos\theta$ in Tables \ref{table_coeffs_yz} and \ref{table_coeffs_xz}.

\begin{table}
\centering
\begin{tabular}{|c|c|c|c|c|c|c|}
\hline
$\varphi_1$ & d &  $A_0$ & $A_1$ & $A_2$ & $A_3$ & $A_4$ \\
\hline
0 (``$yz$ series")&    10 & -0.0679461  &  -1.3985e-4 & -1.07477e-6 &  3.47174e-7 & -1.68858e-7 \\
0.175 &                10 & -0.067933   &  -1.3891e-4 & -1.45311e-5 &  -1.39774e-6 & 4.15175e-7 \\
1.57 (``$xz$ series")& 10 & -0.0676704  &  -1.3787e-4 & -2.74078e-4 &  -1.55601e-6 & -2.17319e-6 \\
\hline
\end{tabular}
\caption{
Table of coefficients for curve fits of the form 
$\BENorm = A_0 + A_1\cos\theta + A_2\cos^2\theta + A_3\cos^3\theta + A_4\cos^4\theta$,
for a separation of $10m$.  
(Here and below, we report several significant digits for the purposes of comparison, however due
to variations resulting from excision region size and other effects, one would rightly regard only 
the first two digits as significant.)
This shows that terms higher in
order than $\cos^2\theta$ do not contribute significantly.  Because
of this, we do not include powers higher than two in the trigonometric basis functions.
Also, given the considerations due to the mass quadrupole moment in Eq. (\ref{mass_quad}),
{\it all subsequent curve fits in this paper use the form 
$\BENorm = B_0 + B_1\cos(\theta) + B_2\sin^2\theta$.}  That is, the second order term will be taken as proportional to
$\sin^2\theta$, not $\cos^2\theta$.  This results in an offset of the total binding energy ($A_0$).
}
\label{table_coeffs}
\end{table}

\begin{figure}
\centering
\includegraphics[width=4in]{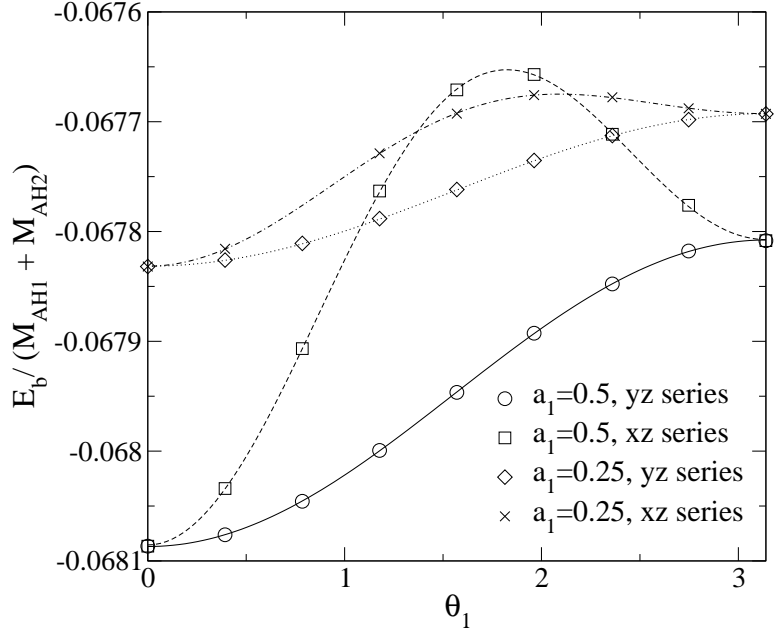}
\vspace{-0.5cm}
\caption{
The effects of varying the spin magnitude of one of the holes.
Symbols denote data points, lines denote curve fits.  Notably, the
fit for the $xz$ series with $a_1=0.5$ is 
$\BENorm = -0.0679464 -1.39041\times 10^{-4}\cos\theta + 2.76249\times 10^{-4}\sin^2\theta$, 
while for the $xz$ series with $a_1=0.25$ it is 
$\BENorm = -0.0677621 -6.94328\times 10^{-5}\cos\theta + 7.00104\times 10^{-5}\sin^2\theta$.  
Thus reducing spin $a_1$ from 0.5 to
0.25 results in reduction of the $\cos\theta$ term by a factor of
two, while the $\sin^2\theta$ term is reduced by nearly a factor
of four.  
} 
\label{cutspininhalf} \end{figure}

We tested the spin-squared dependence of the $\sin^2\theta$ term
by two methods. In one we considered $a=a_1=0.25m$ for the hole at
$x=-5m$, which was then tested in an abbreviated $yz$ series, while
$a_2$ held at $a_2=0.5$ along the positive $z$ axis for the hole
at $x=+5m$. Figure \ref{cutspininhalf} shows the result; the effect
from the quadrupole term is quadratic in the reduced spin (its
amplitude is reduced by a factor of four), while the Wald spin-spin
interaction is linear in the reduced spin and its amplitude is
reduced by a factor of two.  To further test the quadrupole dependence
we considered rotating the spin of the black hole at $x=-5m$ in a
plane turned by $\varphi_1=0.175{\rm rad} \approx 10^\circ$.  The
coefficients of the nonlinear curve fit are listed on the second
line of Table \ref{table_coeffs} and are plotted as interpolating
lines in Figure \ref{be_vs_theta_xz}.  They have the analytically
expected dependence.

\begin{table}
\centering
\begin{tabular}{|c|c|c|c|c|}
\hline
Series & $\sep$ & $B_0$ & $B_1$ & $B_2$  \\
\hline
$yz$   &  6     &    -0.0743885  & -0.000381203  &  4.68096e-6   \\
$yz$   &  7     &    -0.0731073  & -0.000275675  &  3.11827e-6   \\
$yz$   &  8     &    -0.0716993  & -0.000211891  &  3.06277e-6   \\
$yz$   &  10    &    -0.0679473  & -0.000139598  &  1.25165e-6   \\
$yz$   &  12    &    -0.0632223  & -9.97935e-5  &  -3.78722e-7   \\
$yz$   &  14    &    -0.0576223  & -7.85976e-5  &  -8.578e-8   \\
$yz$   &  16    &    -0.0517695  & -6.45692e-5  &  -3.17302e-7   \\
$yz$   &  18    &    -0.0459505  & -5.52391e-5  &  -1.16e-6   \\
\hline
\end{tabular}
\caption{
Table of coefficients for curve fits of the form 
$\BENorm = B_0 + B_1\cos\theta + B_2\sin^2\theta$, for the $yz$ series ($\varphi_1=0$),
 for various BBH separations $\sep$ with both spins $a_1=a_2=0.5$,
using a domain size of $\pm 15m$ and $513^3$ fine grid points.  Notice that,
as expected, $B_2$ is very small for all these fits.
}
\label{table_coeffs_yz}
\end{table}

\begin{table}
\centering
\begin{tabular}{|c|c|c|c|c|}
\hline
Series & $\sep$ & $B_0$ & $B_1$ & $B_2$  \\
\hline
$xz$   &  6     &    -0.0743885  & -0.000381203  &  0.000431581   \\
$xz$   &  7     &    -0.0731073  & -0.000275674  &  0.000357985   \\
$xz$   &  8     &    -0.0716993  & -0.00021189   & 0.000316921   \\
$xz$   &  10    &    -0.0679464  & -0.000139041  &  0.000276249   \\
$xz$   &  12    &    -0.0632223  & -9.97932e-5  &  0.000224196   \\
$xz$   &  14    &    -0.0576223  & -7.85974e-5  &  0.000193869   \\
$xz$   &  16    &    -0.0517695  & -6.45689e-5  &  0.000171335   \\
$xz$   &  18    &    -0.0459505  & -5.52389e-5  &  0.000146205   \\
\hline
\end{tabular}
\caption{
Table of coefficients for curve fits of the form 
$\BENorm = B_0 + B_1\cos\theta + B_2\sin^2\theta$, for the $xz$ series ($\varphi_1=1.57$),
for various BBH separations $\sep$,
using a domain size of $\pm 15M$ and $513^3$ fine grid points.
}
\label{table_coeffs_xz}
\end{table}

\begin{figure}
\centering
\includegraphics[width=4in]{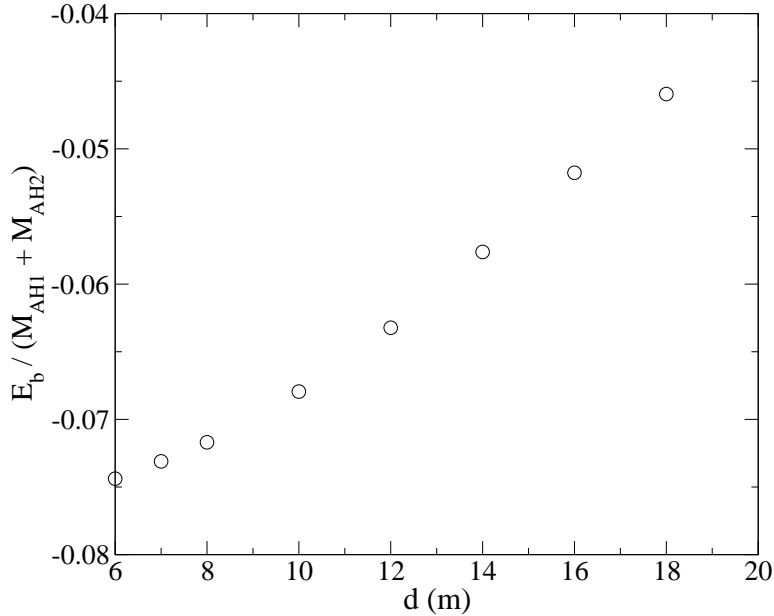}
\caption{
Variation of binding energy vs. BBH separation $\sep$ (holes at $x=\pm \sep/2$), showing the constant term $B_0$ in the curve fits shown
in Table \ref{table_coeffs_yz}.  
The circles are computed using $513^3$ grid points on a domain of $\pm 15m$, 
evaluating the ADM mass on a cube of half-width $12m$.
We note that, at large separations, the curve is roughly linear, in contrast to an
expected behavior of $1/\sep$.  We speculate that this (unphysical) effect is due to the outer 
boundary of the computational domain, in particular the behavior of the ADM mass as estimated at these
rather small outer radii.  However it may simply be the case that the asymptotic behavior only evident
at larger separations.
We expect that future work using our multiresolution code with larger domains
and higher resolutions should produce better agreement with the expected behavior.
}
\label{amp_vs_sep_const}
\end{figure}

\begin{figure}
\centering
\includegraphics[width=4in]{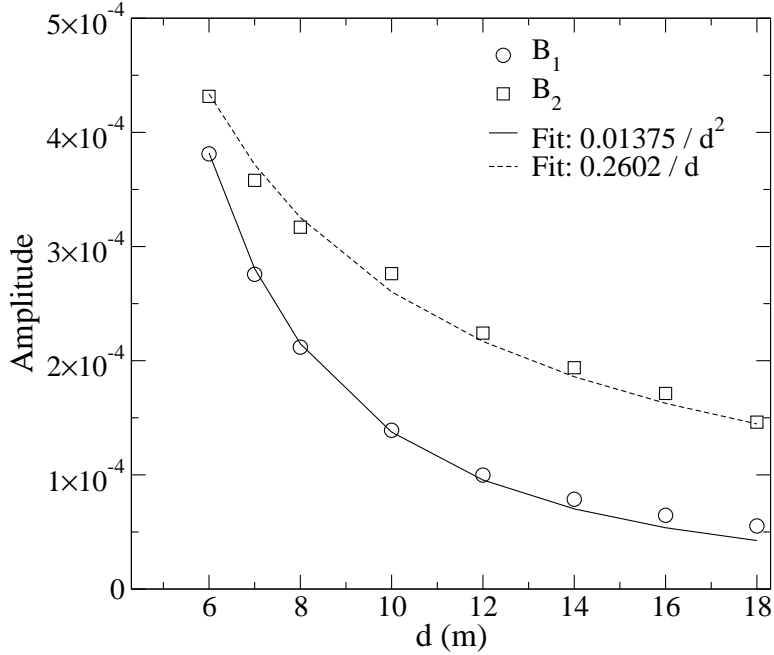}
\vspace{-0.5cm}
\caption{
Amplitude vs. separation for the constants $B_1$ in Tables
\ref{table_coeffs_yz} and \ref{table_coeffs_xz} (circles), and $B_2$
in Table \ref{table_coeffs_xz} (squares).  Rather than seeing the
expected asymptotic $d^{-3}$ fall-off for each amplitude, we find
the cosine term $B_1$ has roughly $d^{-2}$ behavior, whereas the
$\sin^2$ term $B_2$ falls off no faster than $1/d$.  
} 
\label{amp_vs_sep}
\end{figure}

Figures \ref{amp_vs_sep_const} and \ref{amp_vs_sep} show our tests
of  the separation-dependence of the binding energy.  We expect the
constant term $B_0$ to fall off asymptotically as $1/d$, since it
corresponds to the $M/r$ term in the Newtonian limit.  Instead we
find roughly linear behavior at the largest separations we are able
to compute.  The amplitudes $B_1$ and $B_2$ also fall off differently
than expected.  We expect both $B_1$, which corresponds to the
cosine term in (\ref{BEWald}), and $B_2$, which corresponds to the
mass quadrupole term, to scale as $d^{-3}$.  Instead we find that
$B_1$ scales as $d^{-2}$, and that $B_2$ scales no faster than
$d^{-1}$.  Since we expect the constant term $B_0$ to scale as $1/d$
(although, as in Figure \ref{amp_vs_sep_const}, we see that it does
not), dividing the amplitudes $B_1$ and $B_2$ by the constant $B_0$
does not significantly illuminate the results.  These results for
the separation-dependence are likely affected by the outer boundaries
of our computational domain in unphysical ways.  In the future we
hope to repeat these studies with higher resolution and larger
domains, using a multi-resolution (mesh refinement) version of our
code.  For the present, we conducted an additional test to measure
the effects of the outer boundary, namely we looked for unphysical
effects by computing the difference between the ADM mass and the
horizon mass for a {\em single} black hole as we rotated its spin
axis.  The variation we found was on the order of $10^{-6}$, which
would appear as a horizontal line in Figure \ref{be_vs_theta_yz}.
While this provides some assurance that our mass determination
methods are functioning to some level of expectation, this (single-hole)
effect is insufficient to explain the deviation from the expected
results in the binary simulations.  Future refinements and larger
domains will hopefully clarify this issue.

\section{Discussion}
To the extent checked, our computational tests of the spin-spin interaction in the binding
energy of binary black hole configurations, verify 
the angular dependence given by Equation (\ref{BEWald})
\cite{WaldPRD}.  We verified the behavior given by Wald's $-(\vec{S}
\cdot \vec{S}')$ term, but since we always kept one black hole's
spin axis perpendicular to the separation axis, we made no attempt
to observe the $\vec{S}\cdot\hat{n}$ term in Wald's formula.  At
the separations and domains available,  our results did not show
asymptotic $\sep^{-3}$ fall-off with distance.

Additionally, we find an effect due to the mass quadrupole moment
of the black holes.  This results in a higher-order (sine squared)
variation with spin angle than given in Wald's formula.  Thus we combine
these two effects into a general equation (\ref{HVMEq}) for spin-spin interactions
of binary black holes.  However,
in this case as well, the expected asymptotic fall off of $d^{-3}$
was not evident in our solutions.  Future analysis will use a new
multiresolution version of our code to further pursue questions
raised by the results here, including moving to significantly larger
separations to evaluate the asymptotic behavior of the spin-dependent
interactions, and effects due to rotation of {\em both} holes' spin axes.  
We also intend to investigate the spin-orbit coupling
and its bearing on evolutions such as \cite{Brownsville2}.  We are
now beginning exploration of the constrained evolution approach in
spacetimes involving single moving, and multiple interacting black
holes. We find substantial improvement from constraint solving in
every simulation.

\section*{Acknowledgments}
We thank Evan Turner, Chris Hempel and Karl Schultz of the Texas
Advanced Computing Center at the University of Texas, where the
computations were performed.  This work was supported by NSF grant
PHY~0354842, and by NASA grant NNG04GL37G. Portions of this work
were conducted at the Laboratory for High Energy Astrophysics,
NASA/Goddard Space flight Center, Greenbelt Maryland, with support
from the University Space Research Association.


\newpage

\begin{thebibliography}{99}


\bibitem{YP} J.~York and  T.~\ Piran
        ``The Initial Value Problem and Beyond'',
        \textit{Spacetime and Geometry: The Alfred Schild
        Lectures}, R.~Matzner and L.~Shepley Eds.
        University of Texas Press, Austin, Texas. (1982);
        G.~Cook, ``Initial Data for the Two-Body Problem
        of General Relativity'', Ph.D. Dissertation, The University
        of North Carolina at Chapel Hill (1990).


\bibitem{MY} N.~{\'O}.~Murchadha and J.~W.~York, Jr.,
                   {\it Phys.\ Rev.}  {\bf D10}, 428 (1974);
             N.~{\'O}.~Murchadha and J.~W.~York, Jr.,
                   {\it Phys.\ Rev.} {\bf D10}, 437 (1974);
             N.~{\'O}.~Murchadha and J.~W.~York, Jr.,
                   {\it Gen.\ Relativ.\ Gravit.} {\bf 7} 257 (1976).

\bibitem{York} J.~W.~York, Jr., {\it Phys.\ Rev.\ Lett.} {\bf 82},
        1350 (1999).



\bibitem{Mathews} J.~R.~Wilson and G.~J.~Mathews,
    {\it Phys.~Rev.~Lett.} {\bf 75}, 4161 (1995);
    J.~R.~Wilson and G.~J.~Mathews, and P.~Marronetti,
    {\it Phys.~Rev.} {\bf D54}, 1317 (1996)


\bibitem{Bowen+York} J.~Bowen and J.~W.~York,
                     {\it Phys. Rev.} {\bf D21}, 2047 (1980).

\bibitem{WaldPRD} R.~Wald, {\it Phys.\ Rev.} {\bf D6}, 406 (1972).

\bibitem{Choquet} Y. Choquet-Bruhat, {\it Global Solutions of the
Problem of Constraints on a Closed Manifold}, {\it Symp. Math.}{\bf XII} 317
(1973).

\bibitem{Lichnerowicz} A. Lichnerowicz, ``L'Integration des Equations de la
Gravitation Relativiste et le Probleme des n Corps, {\it J. Math. Pures et
Appl.}
{\bf 23} 37-63 (1944).

\bibitem{Hawley+Matzner}Scott H. Hawley and Richard A. Matzner
{\it Class. Quant. Grav.} {\bf21} 805-822 (2004).




\bibitem{ADM} R.~Arnowitt, S.~Deser, and C.~Misner in Witten,
{\it Gravitation,
an Introduction to Current Research} (Wiley, New York 1962).

\bibitem{Cook}  G.~B.~Cook, {\it Phys.\ Rev.}  {\bf D50}, 5025 (1994).

\bibitem{Pfeiffer} H.~P.~Pfeiffer, S.~A.~Teukolsky and G.~B.~Cook
              {\it Phys.\ Rev.} {\bf D62}, 104018 (2000).

\bibitem{Baumgarte}T.~Baumgarte {\it Phys. Rev.} {\bf D62}, 024018 (2000).

\bibitem{GGB1} E.~Gourgoulhon, P.~Grandclement and S.~Bonazzola,
{\it Phys.\ Rev.}  {\bf D65}, 044020 (2002).

\bibitem{GGB2}
P.~Grandclement, E.~Gourgoulhon and S.~Bonazzola,
{\it Phys.\ Rev.} {\bf D65}, 044021 (2002), and references therein.

\bibitem{GGB3} E.~Gourgoulhon, P.~Grandclement and S.~Bonazzola,
{\it Int.\ J.\ Mod.\ Phys.} {\bf A17} 2689--94 (2002).

\bibitem{Cook2} G.~B.~Cook {\it Phys.\ Rev.}  {\bf D65} 084003 (2002).

\bibitem{Pfeiffer2} H.~P.~Pfeiffer, G.~B.~Cook, and S.~A.~Teukolsky
{\it Phys.\ Rev.}  {\bf D66} 024047 (2002).



\bibitem{CookReview} G.~B.~Cook, {\it Living Rev.\ Rel.} {\bf 3}, 5 (2000).




\bibitem{Shoemaker}D.~Shoemaker, M.~F.~Huq and R.~A.~Matzner
{\it Phys. Rev.} {\bf D62}, 124005 (2000).

\bibitem{Huq}M.~F.~Huq, M.~Choptuik and R.~A.~Matzner,
{\it Phys. Rev.} {\bf D66}, 084024 (2002). [arXiv:gr-qc/0002076].


\bibitem{KerrSchild} R.~Kerr and A.~Schild, ``Some Algebraically Degenerate
Solutions of Einstein's Gravitational Field Equations,'' in
{\it Applications of Nonlinear Partial Differential Equations in
Mathematical Physics}, Proc. of Symposia B Applied Math., Vol XVII (1965);
``A New Class of Solutions of the Einstein Field Equations'',
{\it Atti del Congresso Sulla Relitivita Generale: Problemi Dell'Energia
E Onde Gravitazionala} G. Barbera, Ed. (1965).

\bibitem{Brownsville} M.~Campanelli, C.~O.~Lousto, P.~Marronetti, Y.~Zlochower,
{\tt gr-qc/0511048} (2005).

\bibitem{Goddard} J.~G.~Baker, J.~Centrella, D.-I.~Choi, M.~Koppitz, J.~ van Meter,
{\tt gr-qc/0511103} (2005).

\bibitem{Matzner:1999pt}
R.~Matzner, M.~F.~Huq and D.~Shoemaker,
{\it Phys.\ Rev.}  {\bf D59}, 024015 (1999)[arXiv:gr-qc/9805023].

\bibitem{Bonning}Erin Bonning, Pedro Marronetti, David Neilsen and
Richard Matzner {\it Phys. Rev.} {\bf D68} (2003) 044019.

\bibitem{Mattsnotes}Choptuik M W  1999
Lecture notes, Taller de Verano 1999 de FENOMEC:
Numerical Analysis with applications in Theoretical Physics.


\bibitem{BBHGC}The Binary Black Hole Grand Challenge Alliance: G. B. Cook, M. F. Huq, 
S. A. Klasky, M. A. Scheel, A. M. Abrahams, A. Anderson, P. Anninos, T. W. Baumgarte, 
N. T. Bishop, S. R. Brandt, J. C. Browne, K. Camarda, M. W. Choptuik, C. R. Evans, 
L. S. Finn, G. C. Fox, R. Gomez, T. Haupt, L. E. Kidder, P. Laguna, W. Landry, L. Lehner, 
J. Lenaghan, R. L. Marsa, J. Masso, R. A. Matzner, S. Mitra, P. Papadopoulos, 
M. Parashar, L. Rezzolla, M. E. Rupright, F. Saied, P. E. Saylor, E. Seidel, 
S. L. Shapiro, D. Shoemaker, L. Smarr, W. M. Suen, B. Szilagyi, S. A. Teukolsky, 
M. H. P. M. van Putten, P. Walker, J. Winicour, J. W. York Jr., 
``Boosted three-dimensional black-hole evolutions with singularity excision",
{\it Phys. Rev. Lett.} {\bf 80} 2512-2516 (1998) [gr-qc/9711078].

\bibitem{brown}J. David Brown, Lisa L. Lowe, ``Multigrid elliptic equation solver with adaptive mesh refinement",
{\it J. Comput. Phys.} {\bf 209} 582-598 (2005)  [gr-qc/0411112].


\bibitem{KipMoments}K. S. Thorne, ``Multipole expansions of gravitational
radiation"
{\it Rev Mod Phys} {\bf 52} 299 (1980).

\bibitem{walterHernandez}Walter C. Hernandez, ``Material Sources
for the Kerr Metric", {\it Phys Rev} {\bf 159} 1070 (1967)






\bibitem{MTW} C.~W.~Misner, K.~S.~Thorne, and J.~A.~Wheeler, {\it
Gravitation}
                   (W.H. Freeman, New York, 1970).

\bibitem{Wald} R.~ Wald, {\it General Relativity}
        (University of Chicago Press, Chicago, 1984).


\bibitem{Marronetti:2000rw}
P.~Marronetti, M.~Huq, P.~Laguna, L.~Lehner, R.~Matzner and
D.~Shoemaker, {\it Phys.\ Rev.}  {\bf D62}, 024017 (2000) [gr-qc/0001077].

\bibitem{Marronetti}P.~Marronetti and R.~A.~Matzner
{\it Phys. Rev. Lett.} {\bf 85} 5500 (2000).

\bibitem{Dain} S.~Dain {\it Phys. Rev.}{\bf D66} 084019 (2002).

\bibitem{note4} For consistency of notation we
use the superscript ``ADM" to decorate the angular momentum
$J^{\ADM}_{ab}$. We are not certain where the formula (\ref{eq:adm_ang_mom})
was first written. It does {\it not} appear in~\cite{ADM}.
 Expressions equivalent to it appear in~\cite{Weinberg}
and in~\cite{MTW}, and were promulgated in notes by Misner (unpublished) in
the
mid-60s, arising from the long history of conservation law/pseudo-tensor
studies in General Relativity. The form here was taken from~\cite{Wald}; see
also the very thorough development in~\cite{Brown}.


\bibitem{Weinberg} S.~Weinberg, {\it Gravitation and Cosmology}
   (Wiley and Sons, New York, 1972)

\bibitem{Brown} J.~D.~Brown and J.~W.~York, {\it Phys. Rev.} {\bf D47},
        1407 (1993).


\bibitem{Brill} D.~Brill and R.~W.~Lindquist
            {\it Phys. Rev.} {\bf 131} 471 (1963).

\bibitem{Misner} Misner, C.~W., ``The Method of Images in Geometrostatics'', 
       {\it Ann. Phys.} (N. Y.), {\bf 24}, 102, (1963).

\bibitem{cadez} 
   This was apparently first noticed as a  computational result; see
   A. \v Cade\v z, {\it Ann. Phys. (NY)} {\bf 83} 449 (1974). \v Cade\v z
   considered both the Brill-Lindquist~\cite{Brill} and
   Misner~\cite{Misner} data. For Brill-Lindquist data the apparent
horizon areas are easy to compute to $\rmO(m m'/\sep)$, because the lowest
order effect of the distant hole on the local one is a constant, isotropic
addition to the local value of the conformal factor.


\bibitem{NumRecepies} W.~H.~Press, S.~A.~Teukolsky, W.~T.~Vetterling,
and B.~P.~Flannery,
{\it Numerical Recipies in Fortran, Second Edition} (Cambridge University
Press, Cambridge, 1992).


\bibitem{Pfeiffer:2003} H.~Pfeiffer L.~Kidder, M.~Scheel and S.~A.~Teukolsky,
{\it Comput. Phys. Commun.} {\bf 152} 253 (2003).

\bibitem{Brady:1998}P.~Brady, J.~Creighton and K.~S.~Thorne,
{\it Phys. Rev.} {\bf D58}, 061501 (1998).

\bibitem{Klasky} S.~Klasky, PhD Dissertation, University of Texas at Austin,
        (1994).


\bibitem{gr-qc/0002076}
Mijan F. Huq, Matthew W. Choptuik and Richard A. Matzner,
``Locating boosted Kerr and Schwarzschild apparent horizons,"
{\it Phys.\ Rev.\ } {\bf D66}, ~084024~(2002).



\bibitem{eric} S. Bonazzola, E. Gourgoulhon, P. Grandcl\'{e}ment and J.
Novak,
``A constrained scheme for Einstein equations based on Dirac gauge
and spherical coordinates,"
{\it Phys. Rev.} {\bf D70}, 104007 (2004)
.
\bibitem{projection}G. Toth ``The ${\bf \nabla \cdot}$ {{\bf B}} $= 0$
constraint in shock-capturing
magnetohydrodynamics codes,"
{\it Jour. Comp. Phys.} {\bf 161}, 605 (2002).

\bibitem{Holst:2004}
Michael Holst, Lee Lindblom, Robert Owen, Harald P. Pfeiffer, Mark A. Scheel
and 
Lawrence E. Kidder,
``Optimal constraint projection for hyperbolic evolution systems,"
gr-qc/0407011 (2004).

\bibitem{mattDiss}Matthew Anderson, ``Constrained evolution in numerical
relativity,"
Ph.D. dissertation, The University of Texas at Austin (2004).

\bibitem{ThornburgAHFinder} J. Thornburg, ``A Fast Apparent-Horizon Finder for 3-Dimensional Cartesian Grids in Numerical Relativity," {\it Class. Quant. Grav.} {\bf21} 743-766 (2004).
 
\bibitem{cactus-grid}  G. Allen, W. Benger, T. Dramlitsch, T. Goodale, H.-C. Hege, G. Lanfermann, A. Merzky, T. Radke, 
     and E. Seidel, in {\it Euro-Par 2001: Parallel Processing, Proceedings of the 7th International Euro-Par Conference},
     edited by R. Sakellariou, J. Keane, J. Gurd, and L. Freeman (Springer, 2001). 

\bibitem{cactus-tools} G. Allen, W. Benger, T. Goodale, H.-C. Hege, G. Lanfermann, A. Merzky, T. Radke, E. Seidel, and J. Shalf, 
        {\it Cluster Computing} {\bf 4}, 179 (2001). 

\bibitem{cactus-review} B. Talbot, S. Zhou, and G. Higgins (2000), {\tt http://sdcd.gsfc.nasa.gov/ESS/esmf tasc/Files/Cactus b.html.}

\bibitem{cactus-webpages} The Cactus Team, The Cactus computational toolkit, {\tt http://www.cactuscode.org}.

\bibitem{Goodale02a} T. Goodale, G. Allen, G. Lanfermann, J. Mass  o, T. Radke, E. Seidel, and J. Shalf, 
   in {\it Vector and Parallel Processing  VECPAR2002, 5th International Conference, Lecture Notes in Computer Science}
      Springer, Berlin (2002). 

\bibitem{Brownsville2} M.~Campanelli, C.~O.~Lousto, P.~Marronetti, Y.~Zlochower,
{\tt gr-qc/0604012} (2006).

\end{thebibliography}
\end{document}